# A Systematic Literature Review on relationship between agile methods and Open Source Software Development methodology

Taghi Javdani Gandomani[1], Hazura Zulzalil[2], Abul Azim Abd Ghani[3], Abu Bakar MD Sultan[4]

**Abstract** – *Agile software development methods (ASD) and open source software development methods (OSSD) are two different approaches which were introduced in last decade and both of them have their fanatical advocators. Yet, it seems that relation and interface between ASD and OSSD is a fertile area and few rigorous studies have been done in this matter. Major goal of this study was assessment of the relation and integration of ASD and OSSD. Analyzing of collected data shows that ASD and OSSD are able to support each other. Some practices in one of them are useful in the other. Another finding is that however there are some case studies using ASD and OSSD simultaneously, but there is not enough evidence about comprehensive integration of them.*

## I.   Introduction

Agile software development methodology (ASD) which was introduced formally within last decades is a reaction to traditional methods (e.g. waterfall) which also are recognized as heavy-weight methods [1]. Since technology and industry grow too fast, requirements changes rapidly and so, new innovation of software development was introduced [2]. Four agile principles introduced as agile manifesto [3], have attracted many software producers and software engineers to migrate from plan-based software development methodology to agile methods [4]-[5]-[6]-[7]-[8] . Also, focusing on user requirements and embracing of changes during development cycle causing customers to accept these methods as well as welcoming. Open Source Software Development (OSSD) is another famous approach which has been used increasingly in last decade. OSSD relies on skilled volunteers and their experiences in distributed teams [9]. However there is no specific definition, OSSD is defined as a revolutionary software process allowing source codes to be redistributed and modified freely and encourage developing software in a collaborating environment. This approach is also increasingly used in many projects and organizations [10]. Both ASD and OSSD provide many benefits and facilities that could be effective in software production. From the early years, the relationship between these approaches was studied [11]-[12].
Aim of this study is assessing of the relationship between ASD and OSSD, and if so, extent of the relation and probably integration of them by doing a systematic Literature Review (SLR) on simple and clear question in this regard.

In the next section we describe our method for conducting review, in section III results will be shown, in section IV we answer our questions and finally in section V, conclusion will be presented.

## II.   Method

The goal of this study is conducting SLR to assess the relationship between ASD and OSSD. We used guidelines proposed by Kichenham [13] for performing our study. The main steps are explained in the next parts of this section.

### II.1.   Research Questions

The addressed questions in this study are:

RQ1: Could ASD and OSSD have any relationship?
RQ2: Are practices of one of them applicable in the second?
RQ3: Can they integrate with each other?

As regards to RQ1, it was important for us to know any possible relation between these approaches. Our aim was not to address adequately relation of them; we were only looking for finding any relation. If so, then we could focus on next questions.

As regards to RQ2, we were looking for finding any practices or rules in one them which could be applicable by the second. For this question, we focused on software engineering concepts in either method.

As regards to RQ3, We were looking for researches and case studies that had reported in combining, integrating or collaborating of both methods in software projects or organizations. We were looking for feasibility of

Taghi Javdani Gandomani[1], Hazura Zulzalil[2], Abul Azim Abd Ghani[3], Abu Bakar Md Sultan[4]

application of both methods simultaneously, even in many specific practices.

### II.2.    Research Process

Our search process for review was based on online searching in famous online databases which are addressed as table I. Since these databases cover almost all major journals and conference proceedings, manually review of journal was not required. Review has been carried on by mean of search facilities in these databases and using appropriate logical expressions. In first stage, our focus was on title and abstract of articles found in search process and select appropriate and relevant studies. If there was any doubt, our decision was based on reviewing it at one glance. If not sure for choosing it, in next step we asked an expert to help us in decision making. Our final step was direct contact to authors, which was not used in selection process.

TABLE I
STUDIES RESOURCE

| Source | Address |
| --- | --- |
| Scopus | www.scopus.com |
| IEEE Xplore | ieeexplore.ieee.org |
| ACM Digital Library | Portal.acm.org |
| Springer Link | www.springerlink.com |
| Tailor and Francis | www.tandfonline.com |
| Science Direct | www.sciencedirect.com |

### II.3.    Inclusion and exclusion criteria

Papers which only had focused on one approach and did not discuss the other were ignored. There were some papers which were relevant to our study indirectly, but, in our defined process, we could not find them. This is not a threat for our review, because all appropriate papers were included and our review covered enough direct studies in this research.

### II.4.    Quality Assessment

For assessing studies we defined the following questions:
QA1. Does study agree with existence of any relation between ASD and OSSD?

QA2. Does study report any similar practice in both methods?

QA3. Does study report successful use of both methods simultaneously?

QA4. Is there any practice in one of them which is useful and helpful in the second?

QA5. Is there any successful case study in integrating them?

QA6. Does study agree with feasibility of integration or collaboration of ASD and OSSD?

We scored questions as bellow:

QA1. Y (Yes) study explicitly agrees with existence of any relationship; P (Partially) study implicitly agrees and N (No) study disagrees with existence of any relation.
QA2. Y, the authors address one or more similar practices; P, some of the ones practices could be tailored and customized in the second and N, there is no similar and adaptable practices in them.

QA3. Y, Authors address successful case study using both ASD and OSSD simultaneously; P, Authors address some case studies which use some of the ASD practices and some of the OSSD practices simultaneously and N, there is no case study of application of ASD and OSSD simultaneously.

QA4. Y, the authors report any useful practices of one that is applicable and helpful in the second; P, customized and tailored practices of each one, could be benefit in the second and N, there is no practice of each one, usable in the second.

QA5. Y, study addresses successful case study of integration of ASD and OSSD; P, study addresses case study of integration ASD and OSSD which is partly successful and N, there is no successful case study of integration of ASD and OSSD.

QA6. Y, study agrees with feasibility of integration of ASD and OSSD; P, study partly agrees (or implicitly) with integration ASD and OSSD and N, study rejects feasibility of integration between ASD and OSSD or has no idea about it.

We defined Y=1, P=0.5 and N=0 or Unknown where information is not clearly specified. All authors assessed every article and if there is no agreement in scoring, we discussed enough to reach agreement. For unknown questions we should ask study's authors via email and re-score question based on received answers.

### II.5.    Data Collection

These data were extracted from each article:

- The full source and references
- The author(s) information and details
- Research issues
- Main ideas and our questions related information and candidate answers

All articles were reviewed and data was extracted by one person and checked by another. This idea was chosen for better consistency in reviewing all papers and improving quality of review. In any disagreement, authors discussed to reach to an agreement.

Taghi Javdani Gandomani[1], Hazura Zulzalil[2], Abul Azim Abd Ghani[3], Abu Bakar Md Sultan[4]

### II.6.  Data Analysis

Our collect data was organized to address:

- Whether study agrees with existence of any relationship between ASD and OSSD or not? (Addressing RQ1)
- Whether study mentions similar practice/concept in either methods or no? (Addressing RQ2)
- Whether study reports simultaneously application of two methods or no? (Addressing RQ2, RQ3)
- Whether study addresses any useful practice of one method that is applicable in the other or not? (Addressing RQ2 and RQ3)
- Whether study reports any successful integration of two methods or not? (Addressing RQ3)
- Whether authors believe that ASD and OSSD integration or collaboration is feasible or no? (Addressing QR3)

## III.  Results

In this section we explain results of our review.

### III.1.  Search Results

Table II shows the results of our selection procedure. In this table, results of searching in all databases are provided, but, some of the studies were repeated in more than one online database, so, final number of unique studies selected for our review was distinguished after elimination of repeated articles. Final selected studies are listed in table III.

### III.2.  Quality evaluation of studies

During this phase, we found out that some of the selected articles however claimed to be related to both ASD and OSSD, but, they do not provide any valuable information to our research, so, we decided to delete them from scope of our study. List of this study and main reason for eliminating of them is provided in table IV.

Assessment of each study was done by means of criteria explained in section 2.4 and the scores for each of them are shown in table V.

### III.3.  Quality factors

For assessing results of our quality questions, we use average of total scores. This average is useful for some questions, but it is not useful for some other. For instance, we cannot answer the question about possibility of integration with average of scores because of the nature of the question; instead, we use negative ideas for rejecting possibility.

## IV.  Discussion

In this part, the answers to our study questions will be discussed.

### IV.1.  Relation between ASD and OSSD: yes or no?

Most of the articles agree that there are relationship between ASD and OSSD. By reviewing them, it seems that this relation is mostly in how to manage ASD and OSSD project. Also in some studies authors claim that OSSD is one type of ASD [11]-[14]. Hence, our research results support our first question strongly. It seems that most of the authors agree with the perception of relationship between two approaches. This relation sometimes could be found in comparing their features [15]-[16].

### IV.2.  Support Practices

17 studies strongly believe that ASD and OSSD have similar practices. These similar practices mainly return to principles of ASD and OSSD; e.g. both of them rely on self-organized teams and shared goals in team inputs [17]. Of course they have different management [15], but, teams are based on creativity of individuals. Another main common issue is incremental development. However there is different point of view about software development in these methods, but, multi releases is a common concept in both of them.

Nevertheless, only one half of studies present case studies that have used both methodologies simultaneously. In two studies, using TDD in an open source project was not only successful but also caused better code quality [18]-[19]. Three other studies were reported about one European project, 'PyPy'. However they had focused on different view, but, all had talked about successfully using ASD and OSSD simultaneously [20]-[21]-[22]. Eclipse was another project which has used ASD and OSSD simultaneously [23]. Interestingly, all of the papers agree that ASD and OSSD are able to support each other, even at least in some specific practices or areas [9] . One study explained how they used agile practices in a safety-critical open source project [9]. Authors had claimed that using these approaches together afford benefits to both of them. Another author had claimed that agile method helped his team to track the progress in open source project [24]. In some other studies, authors had tried to use concepts of one methodology to another, which were out of our scope [25]-[26].

In sum up, our research shows that ASD and OSSD can help each other and collaborate in some practices.

Taghi Javdani Gandomani[1], Hazura Zulzalil[2], Abul Azim Abd Ghani[3], Abu Bakar Md Sultan[4]

TABLE II
RESULTS OF STUDY SELECTION PROCEDURE

| Source | Search Results | Selected Studies |
|---|---|---|
| Scopus | 113 | 18 |
| IEEE Xplore | 70 | 4 |
| ACM Digital Library | 18 | 4 |
| Springer Link | 33 | 9 |
| Science Direct | 4 | 3 |
| Total | - | 38 |
| Repeated articles | | 11 |
| Finally selected articles | | 27 |

### IV.3. Integration of ASD and OSSD

One critical issue in our study was integration of ASD and OSSD. We do not see any claim on comprehensive integration of these methods. Only about a quarter of studies have presented evidence about integration. We were looking for case studies which claim on integration of these methods directly, but we only found that there is collaboration between them and nothing more. It seems that authors were cautious about this matter in their case studies. Most of them implicitly believe that integration of ASD and OSSD is possible, at least in some specific practices or projects [27].Meanwhile some studies did not disagree with possibility of ASD and OSSD integration [9]-[15]-[27]-[28].Also some of them mentioned that adoption is a necessary activity for using ASD and OSSD together [29]. Three studies [23]-[30]-[31], by presenting successful case studies and another one [32], without any case study agreed with possibility of successful integration. So, we have not found any clear case study on successful integration of both methodologies, but it seems that integration in some practices is feasible by doing appropriate adoption.

TABLE III
SELECTED STUDIES FOR CONDUCTING REVIEW

| ID | Title | Author(s) | Main Topic | Year |
|---|---|---|---|---|
| S1 | Open source development and Agile methods | Simmons and Dillon | Relation between ASD and OSSD | 2003 |
| S2 | Is Open Source Software Development Essentially an Agile Method? | Warsta and Abrahamsson | Relation between ASD and OSSD | 2003 |
| S3 | Introducing TDD on a free libre open source software project: a simulation experiment | Turnu et al. | The effects of adopting TDD on our open source | 2004 |
| S4 | Agile Principles and Open Source Software Development: A Theoretical and Empirical Discussion | Koch | Investigation on accordance of ASD and OSSD | 2004 |
| S4 | Agile, open source, distributed, and on-time - Inside the eclipse development process | Gamma | Discuss on Eclipse project | 2005 |
| S5 | In search of the sweet spot: agile open collaborative corporate software development | Theunissen et al. | Combining ASD and OSSD | 2005 |
| S6 | Open source development and Agile methods | Fraser et al. | Strategies, tools, and communities focused on OSSD | 2006 |
| S7 | Open source software in an agile world | B. Düring | A case study in combining ASD and OSSD(PyPy) | 2006 |
| S8 | Sprint driven development: Agile methodologies in a distributed open source project (PyPy) | B. Düring | A case study in combining ASD and OSSD(PyPy) | 2006 |
| S9 | Trouble in paradise: the open source project PyPy, EU-funding and agile practices | Turnu et al. | Study the effects of the adoption of agile practices on OSSD | 2006 |
| S10 | Modeling and simulation of open source development using an agile practice | Porruvecchio et al. | Relation between ASD and OSSD | 2007 |
| S11 | An agile approach for integration of an open source health information system | Avotins et al. | How agile can help OSSD | 2007 |
| S12 | The case for innovative open source development and agile methods | Goth | A case study in combining ASD and OSSD | 2007 |
| S13 | Sprinting toward open source development | Deshpande and Riehle | Investigation on impact of continuous integration of ASD on OSSD | 2008 |
| S14 | Continuous integration in open source software development | Theunissen et al. | Relation between ASD and OSSD | 2008 |
| S15 | Corporate, Agile and Open Source Software development: A witch's brew or an elixir of life? | Adams and Capiluppi | evaluating the impact of sprinting on a Free Software project | 2009 |
| S16 | Bridging the gap between agile and free software approaches: The impact of sprinting | Tsirakidis et al. | Similarity and differences of ASD and OSSD | 2009 |
| S17 | Identification of success and failure factors of two agile software development teams in an open source organization | Wusteman | Adoption of ASD and OSSD | 2009 |
| S18 | OJAX: A case study in agile Web 2.0 open source development | Lavazza et al. | A case study in using SCRUM for the development of an OSSD | 2010 |
| S19 | Applying SCRUM in an OSS Development Process: An Empirical Evaluation | Corbucci and Goldman | Identify communication issues encountered in ASD and OSSD | 2010 |
| S20 | Open Source and Agile Methods: Two Worlds Closer than It Seems | K. Gray et al. | A case study in using agile methods in OSSD | 2011 |
| S21 | Agile methods for open source safety-critical software | Okoli and Carillo | Comparing OSSD with ASD and Disciplined methods | 2011 |
| S22 | The best of adaptive and predictive methodologies: Open source software development, a balance between agility and discipline | Magdaleno et al. | A review study on relation of famous software development methods. | 2012 |
| S23 | Reconciling software development models: A quasi-systematic review | Simmons and Dillon | Relation between ASD and OSSD | 2003 |

| Title | Author(s) | Year | Reason for Rejection |
|---|---|---|---|
| *AOSTA: Agile Open Source Tools Academy* | Wild | 2006 | It is a workshop report about a tool and cannot be valid for our study. |
| *System development methodologies: A knowledge perspective* | Kerley et al. | 2006 | It discusses on SD methodologies in point of view of knowledge without concern about their relation to each other. |
| *Detecting agility of open source projects through developer engagement* | Adams et al. | 2008 | It focuses on determination of agility in open source projects. Finding agility of OSS is its main aim and not relation or even combination of ASD and OSSD. |
| *Detecting agility of open source projects through developer engagement* | Dimitropoulos | 2009 | This paper cannot answer none of our questions, it only explain that open source can play a role as agility enabler in projects. |

# V.  Conclusion

Software engineers in last decade have been interested in agile methodology and open source software development. Both of them present some new features and they seem beneficial for better and faster software development. By doing an SLR we were looking for relationship between ASD and OSSD. Fortunately our study shows that both ASD and OSSD can help each other and collaborate in doing software projects by sharing their practices. There are enough evidences that agile and open source practices can support each other, mainly because of some of their common concepts and principles. Also, however, there are a few successful experiences on integration of ASD and OSSD, but, most of the studies are optimistic in possibility of their integration, but there is no empirical successful case study for supporting this idea in software producing industry.

TABLE V
QUALITY EVALUATION

| Source | QA1 | QA2 | QA3 | QA4 | QA5 | QA6 |
|---|---|---|---|---|---|---|
| *1* | Y | N | P | Y | N | P |
| *2* | Y | Y | N | Y | N | P |
| *3* | Y | N | Y | Y | P | P |
| *4* | Y | P | N | Y | N | P |
| *5* | Y | Y | Y | Y | P | Y |
| *6* | Y | Y | N | Y | N | Y |
| *7* | Y | Y | N | Y | N | P |
| *8* | Y | Y | Y | Y | P | P |
| *9* | Y | Y | Y | Y | P | P |
| *10* | Y | Y | Y | Y | P | P |
| *11* | Y | Y | P | Y | N | P |
| *12* | Y | Y | N | Y | N | P |
| *13* | Y | Y | Y | Y | P | P |
| *14* | Y | Y | N | Y | N | N |
| *15* | Y | Y | Y | Y | N | P |
| *16* | Y | N | N | Y | N | N |
| *17* | Y | Y | P | Y | P | P |
| *18* | Y | Y | Y | Y | Y | Y |
| *19* | Y | Y | Y | Y | Y | Y |
| *20* | Y | Y | N | Y | N | P |
| *21* | P | P | Y | Y | P | N |
| *22* | P | P | N | Y | N | N |
| *23* | Y | Y | N | Y | N | Y |
| *Average* | 0.96 | 0.80 | 0.50 | 1.00 | 26.1 | 0.52 |

Taghi Javdani Gandomani[1], Hazura Zulzalil[2], Abul Azim Abd Ghani[3], Abu Bakar Md Sultan[4]

## Authors' information

1,2,3,4 Information System Deptartment
Computer Science and Information Technology
University Putra Malaysia

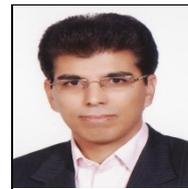

**Taghi Javdani Gandomani** with 10 years experience in both industry and academic research in software methodologies and project management and with the other authors is working in Software Engineering Laboratory in University Putra Malaysia (UPM).
They work on different research in SE, such as software metrics, software quality, software modeling and software methodologies.
Mr. T. Javdani works on agile software methods and he is PhD student in software engineering in University Putra Malaysia.

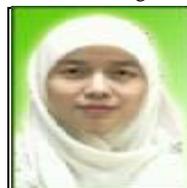

Hazura Zulzalil holds a Ph.D. from University Putra Malaysia. Currently, she is a senior lecturer at the Faculty of Computer Science and Information Technology, University Putra Malaysia. Her research interests are software metrics, software quality and software product evaluation.

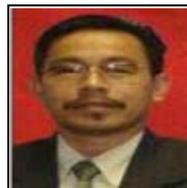

Abdul Azim Abd Ghani obtained his Ph.D. from University of Strathclyde. Currently, he is a Professor at the Faculty of Computer Science and Information Technology, University Putra Malaysia. His research interests are software engineering, software measurement, software quality, and security in computing.

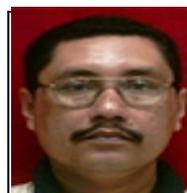

Abu Bakar holds a Ph.D. from University Putra Malaysia. Currently, he is an Associate Professor and the Head of Information Systems Dept., Faculty of Computer Science and Information Technology, University Putra Malaysia. His fields of expertise are Metaheuristic and Evolutionary Computing.
This team, with a lot number of other researchers works on hot issue of software engineering. Many journal and conference papers are results of this team outcome.